\documentclass[runningheads]{llncs}
\usepackage{graphicx}
\usepackage{algorithmic}
\usepackage{graphicx}
\usepackage{textcomp}
\usepackage[dvipsnames]{xcolor}
\usepackage{subcaption}

\usepackage{color}
\usepackage{epsfig}
\usepackage{graphicx}

\usepackage{adjustbox}
\usepackage{array}
\usepackage{booktabs}
\usepackage{multirow}

\usepackage{bm}
\usepackage{nicefrac}
\usepackage{microtype}

\usepackage{changepage}
\usepackage{extramarks}
\usepackage{fancyhdr}
\usepackage{lastpage}
\usepackage{setspace}
\usepackage{soul}
\usepackage{xspace}

\usepackage{marvosym}
\usepackage[hidelinks,colorlinks,citecolor=BlueViolet,linkcolor=BlueViolet]{hyperref}

\begin{document}
\title{Federated Contrastive Learning for Volumetric Medical Image Segmentation\thanks{This work was supported in part by NSF IIS-2027546.}}

\titlerunning{Federated Contrastive Learning for Medical Image Segmentation}

%
\author{Yawen Wu\inst{1}$^{(\textrm{\Letter})}$
\and
Dewen Zeng\inst{2}
\and
Zhepeng Wang\inst{1}
\and
Yiyu Shi\inst{2}
\and
Jingtong Hu\inst{1}
}
%
\authorrunning{Y. Wu et al.}
%
\institute{University of Pittsburgh, Pittsburgh PA 15260, USA \\\email{\{yawen.wu, zhepeng.wang, jthu\}@pitt.edu}\and
University of Notre Dame, Notre Dame IN 46556, USA 
\\\email{\{dzeng2, yshi4\}@nd.edu}
}

\maketitle              
\begin{abstract}

Supervised deep learning needs a large amount of labeled data to achieve high performance. However, in medical imaging analysis, each site may only have a limited amount of data and labels, which makes learning ineffective. Federated learning (FL) can help in this regard by learning a shared model while keeping training data local for privacy. 
Traditional FL requires fully-labeled data for training, which is inconvenient or sometimes infeasible to obtain due to high labeling cost and the requirement of expertise. Contrastive learning (CL), as a self-supervised learning approach, can effectively learn from unlabeled data to pre-train a neural network encoder, followed by fine-tuning for downstream tasks with limited annotations. However, when adopting CL in FL, the limited data diversity on each client makes federated contrastive learning (FCL) ineffective.
In this work, we propose an FCL framework for volumetric medical image segmentation with limited annotations. More specifically, we exchange the features in the FCL pre-training process such that diverse contrastive data are provided to each site for effective local CL while keeping raw data private. Based on the exchanged features, global structural matching further leverages the structural similarity to align local features to the remote ones such that a unified feature space can be learned among different sites. 
Experiments on a cardiac MRI dataset show the proposed framework substantially improves the segmentation performance compared with state-of-the-art techniques.

\end{abstract}

\section{Introduction}

Deep learning (DL) provides state-of-the-art medical image segmentation performance by learning from large-scale labeled datasets \cite{ronneberger2015u,milletari2016v,xu2019whole,dong2017automatic}, without which the performance of DL will significantly degrade \cite{kairouz2019advances}.
However, medical data exist in isolated medical centers and hospitals \cite{yang2019federated}, and combining a large dataset consisting of very sensitive and private medical data in a single location is impractical and even illegal.
It requires multiple medical institutions to share medical patient data such as medical images, which is constrained by the Health Insurance Portability and Accountability Act (HIPAA) \cite{kairouz2019advances} and EU
General Data Protection Regulation (GDPR) \cite{truong2020privacy}.
Federated learning (FL) is an effective machine learning approach in which distributed clients (i.e. individual medical institutions) collaboratively learn a shared model while keeping private raw data local \cite{rieke2020future,sheller2018multi,sheller2020federated,dou2021federated}.
By applying FL to medical image segmentation, an accurate model can be collaboratively learned and data is kept local for privacy.

Existing FL approaches use supervised learning on each client and require that all data are labeled. 
However, annotating all the medical images is usually unrealistic due to the high labeling cost and requirement of expertise.
The deficiency of labels makes supervised FL impractical.
Self-supervised learning can address this challenge by pre-training a neural network encoder with unlabeled data, followed by fine-tuning for a downstream task with limited labels. Contrastive learning (CL), a variant of the self-supervised learning approach, can effectively learn high-quality image representations.
By integrating CL to FL as federated contrastive learning (FCL), clients can learn models by first collaboratively learning a shared image-level representation. Then the learned model will be fine-tuned by using limited annotations.
Compared with local CL, FCL can learn a better encoder as the initialization for fine-tuning, and provide higher segmentation performance.
In this way, a high-quality model can be learned by using limited annotations while data privacy is preserved.

However, integrating FL with CL to achieve good performance is nontrivial. Simply applying CL to each client and then aggregating the models is not the optimal solution for the following two reasons: 
First, each client only has a small amount of unlabeled data with limited diversity.
Since existing contrastive learning frameworks \cite{chen2020simple,he2020momentum} rely on datasets with diverse data to learn distinctive representations, directly applying CL on each client will result in an inaccurate learned model due to the lack of data diversity.
Second, if each client only focuses on CL on its local data while not considering others' data, each client will have its own feature space based on its raw data and these feature spaces are inconsistent among different clients. 
When aggregating local models, the inconsistent feature space among local models will degrade the performance of the aggregated model.

To address these challenges, we propose a framework consisting of two stages to enable effective FCL for volumetric medical image segmentation with limited annotations.
The first stage is feature exchange (FE), in which each client exchanges the features (i.e. low-dimensional vectors) of its local data with other clients. It provides more diverse data to compare with for better local contrastive learning while avoiding raw data sharing. In the learning process, the improved data diversity in feature space provides more accurate contrastive information in the local learning process on each client and improves the learned representations. 

The second stage is global structural matching (GSM), in which we leverage structural similarity of 3D medical images to align similar features among clients for better FCL. 
The intuition is that the same anatomical region for different subjects has similar content in volumetric medical images such as MRI.
By leveraging the structural similarity across volumetric medical images, GSM aligns the features of local images to the shared features of the same anatomical region from other clients. In this way, the learned representations of local models are more unified among clients and they further improve the global model after model aggregation.
Experimental results show that the proposed approaches substantially improve the segmentation performance over state-of-the-art techniques. 

\section{Background and Related Work}

\textbf{Federated Learning.}
Federated learning (FL) learns a shared model by aggregating locally updated models on clients while keeping raw data accessible on local clients for privacy \cite{mcmahan2017communication,li2020federated,zhao2018federated,li2018federated}. 
In FL, the training data are distributed among clients.
FL is performed round-by-round by repeating the local model learning and model aggregation process until convergence.

The main drawback of these works is that fully labeled data are needed to perform FL, which 
results in high labeling costs.
To solve this problem, 
an FL approach using limited annotations while achieving good performance is needed.

\textbf{Contrastive Learning.} 
Contrastive learning (CL) is a self-supervised approach to learn useful visual representations by using unlabeled data \cite{hadsell2006dimensionality,misra2020self,tian2019contrastive}.
The learned model provides good initialization for fine-tuning on the downstream task with few labels \cite{he2020momentum,chen2020simple,chen2020big,zeng2021positional,wu2021enabling}.
CL performs a proxy task of instance discrimination \cite{wu2018unsupervised,chaitanya2020contrastive,wu2021federated}, which maximizes the similarity of representations from similar pairs and minimizes the similarity of representations from dissimilar pairs \cite{wang2020understanding}.

The main drawback of existing CL approaches is that they are designed for centralized learning on large-scale datasets
with sufficient data diversity.
However, when applying CL to FL on each client, the limited data diversity 
will greatly degrade 
the performance of the learned model.
Therefore, an approach to increase the local data diversity while avoiding raw data sharing for privacy is needed.
Besides, while \cite{chaitanya2020contrastive} leverages structural information in medical images for improving centralized CL, it requires accessing raw images of similar pairs for learning.
Since sharing raw medical images is prohibitive due to privacy,
\cite{chaitanya2020contrastive} cannot be applied to FL.
Therefore, an approach to effectively leverage similar images across clients without sharing raw images is needed.

\textbf{Federated Unsupervised Pre-training.}
Some concurrent works employ federated pre-training on unlabeled data.
\cite{van2020towards} employs autoencoder in FL for pre-training on time-series data, but the more effective contrastive learning for visual tasks is not explored in FL.
FedCA \cite{zhang2020federated} combines contrastive learning with FL.
However, it relies on a shared dataset available on each client, which is impractical for medical images due to privacy concerns.

The proposed work differs from these federated unsupervised pre-training approaches in the following ways. First, we do not share raw data among clients to preserve privacy. Second, we leverage the structural similarity of images across clients to improve the quality of representation learning.

\section{Method}

\begin{figure}[!htb]
	\centering
	\includegraphics[width=\columnwidth]{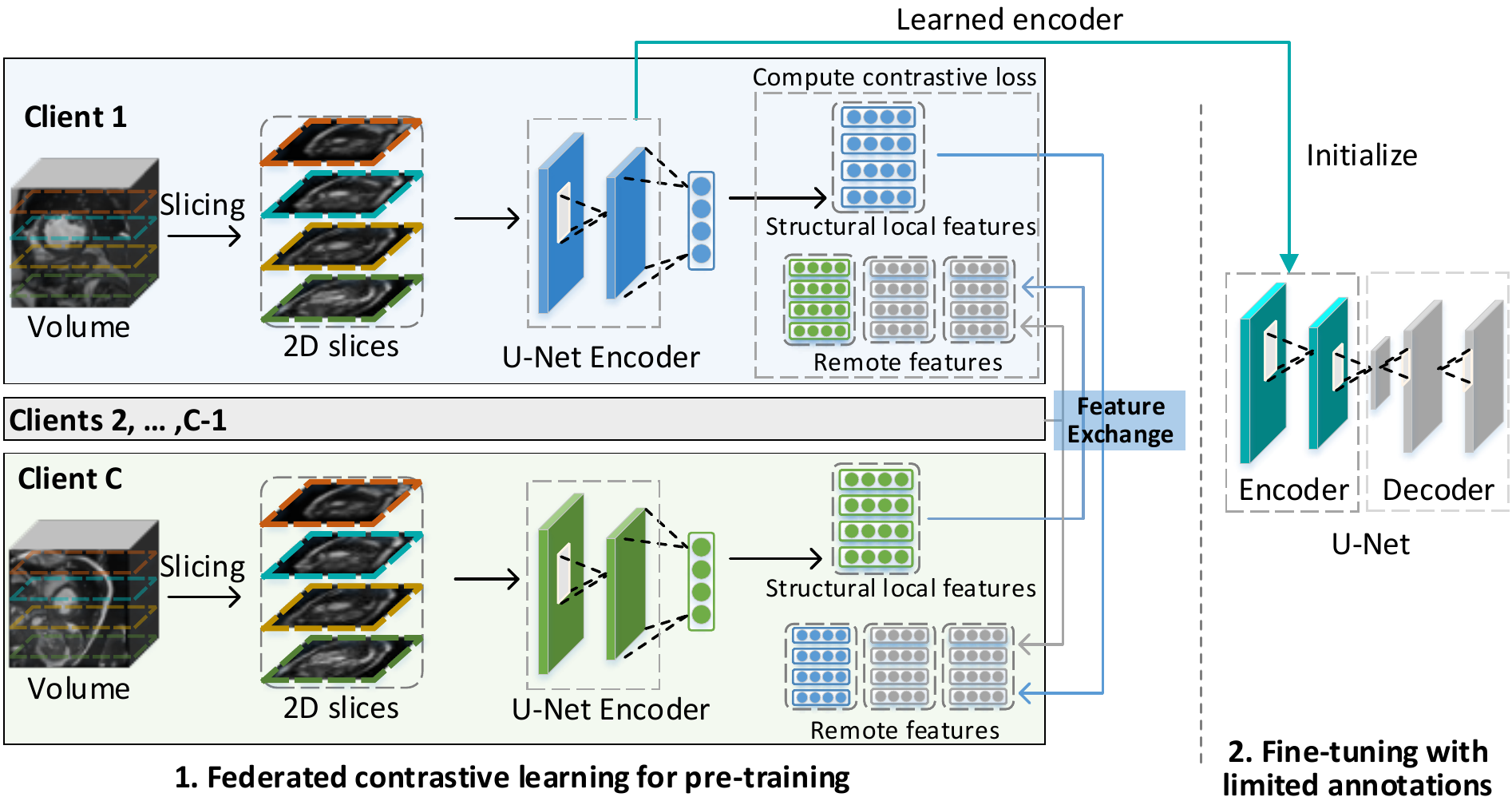}
	\caption{Federated contrastive learning with structural feature exchange for learning the encoder with unlabeled data. 
    Then the learned encoder initializes the encoder in U-Net for fine-tuning with limited annotations.
	}
	\label{fig:overview}
\end{figure}

\textbf{Overview of federated contrastive learning.}
The overview of the proposed FCL process is shown in Fig. \ref{fig:overview}. 
Distributed clients first collaboratively learn a shared encoder by FCL with unlabeled data.
Then the learned encoder initializes the encoder in U-Net \cite{ronneberger2015u} for fine-tuning with limited annotations,
either independently on each client by supervised learning or collaboratively by supervised federated learning. 
Since the supervised fine-tuning can be trivially achieved by using available annotations, in the rest of the paper, we focus on FCL to learn a good encoder as the initialization for fine-tuning.

As shown in Fig. \ref{fig:overview}, in the FCL stage, given a volumetric 3D image on one client, multiple 2D slices are sampled from the volume while keeping structural order along the slicing axis. Then the ordered 2D images are fed into the 2D encoder to generate feature vectors, one vector for each 2D image.

To improve the data diversity in local contrastive learning, one natural way is to share raw images \cite{zhao2018federated}. However, sharing raw medical images is prohibitive due to privacy concerns. 
To solve this problem, the proposed FCL framework exchanges the feature vectors instead of raw images among clients, which can improve the data diversity while preserving privacy.
As shown in Fig. \ref{fig:overview}, client $1$ generates structural local features denoted as blue vectors and shares them with other clients. Meanwhile, client $1$ collects structural features from other clients, 
such as remote features shown in green and gray vectors.
After that, the contrastive loss is computed based on both local and remote features.

\subsection{Contrastive Learning with Feature Exchange}

\begin{figure}[!htb]
	\centering
	\includegraphics[width=\columnwidth]{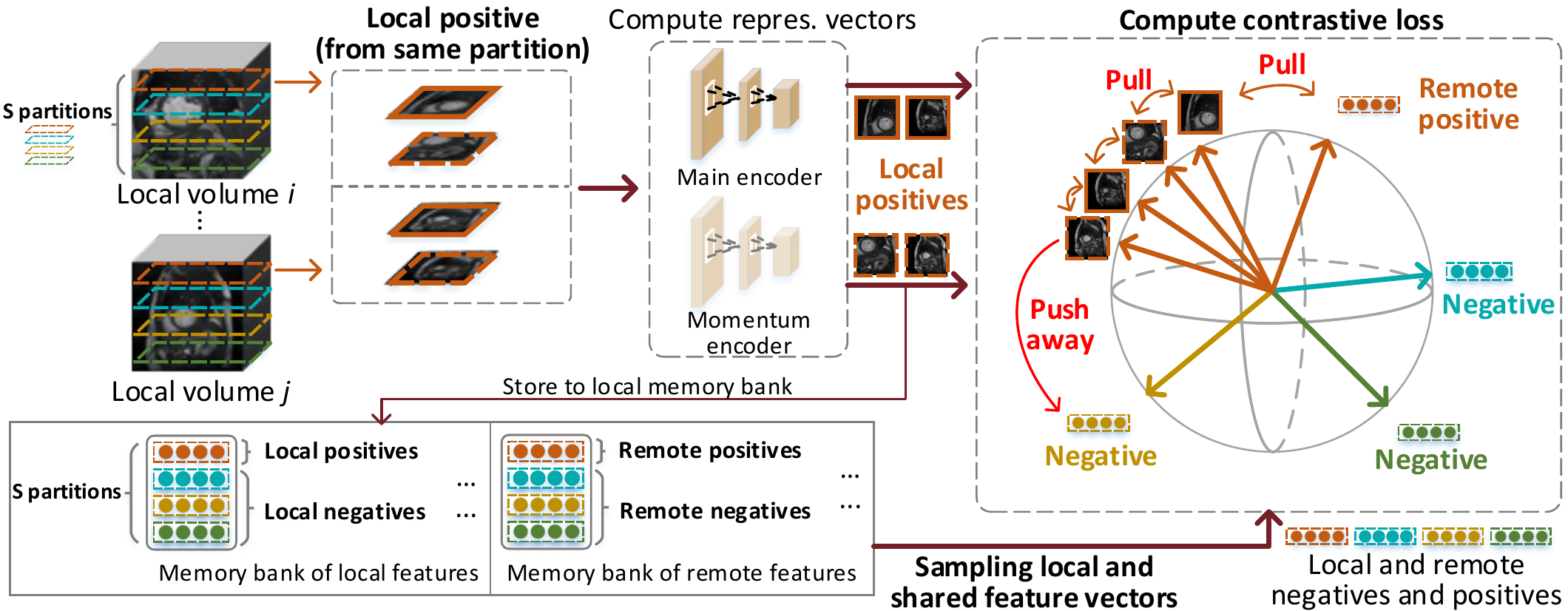}
	\caption{Contrastive learning on one client with exchanged features. The exchanged features consist of remote negatives and remote positives, in which remote negatives improve the local data diversity and remote positives are used for global structural matching to learn a unified feature space among clients.}
	\label{fig:feature_exchange}
\end{figure}

With feature exchange, each client has both remote and local features and is ready to perform local CL in each round of FCL. The remote features provide more diverse features to compare with and improve the learned representations.
As shown in Fig.~\ref{fig:feature_exchange}, we use MoCo \cite{he2020momentum} architecture for local CL since it has a memory bank for negatives, which can leverage local and remote features. 
There are two encoders, including the main encoder and the momentum encoder.
The main encoder will be learned and used as the initialization for fine-tuning, while the momentum encoder is the slowly-evolving version of the main encoder and generates features to contrast with and for sharing. Now the most important steps are to construct negatives and positives from both local and remote features.

\textbf{Negatives from local and remote features.}
Local features are generated by the momentum encoder from local images and used as local negatives.
Each client has a memory bank of local features and a memory bank of remote features.
Let $Q_{l,c}$ be the size-$K$ memory bank of local features on client $c$, which are used as local negatives.
$Q_{l,c}$ is progressively updated by replacing the oldest features with the latest ones.
In each round of FCL, the remote negatives from other clients will be shared with client $c$ to form its aggregated memory bank including local and remote negatives as:
\begin{equation}\label{equ:}
Q = Q_{l,c} \cup \{Q_{l,i}\ |\ 1 \le i \le |C|, i\ne c \}.
\end{equation}
where $C$ is the set of all clients and $Q_{l,i}$ is the local memory bank on client $i$.

Compared with using only local memory bank $Q_{l,c}$, the aggregated memory bank $Q$ provides more data diversity to improve CL. 
However, $Q$ is $|C|$ times the size of the local memory bank $Q_{l,c}$. 
More negatives make CL
more challenging since for one local feature $q$, more negatives need to be simultaneously pushed away from it than when using $Q_{l,c}$, which can result in ineffective learning. 
To solve this problem, instead of using all negatives in $Q$, for each $q$ we sample a size-$K$ (i.e. the same size as $Q_{l,c}$) subset of $Q$ as negatives, which is defined as: 
\begin{equation}\label{equ:contrastive_negatives}
    Q^{\prime}=\{ Q_{i} |\ i \sim \mathcal{U}(|Q|,K) \}.
\end{equation}
where $i \sim \mathcal{U}(|Q|,K)$ means $i$ is a set of indices sampled uniformly from $[|Q|]$.

\textbf{Local positives.}
We leverage the structural similarity in the volumetric medical images to define the local positives, in which the same anatomical region from different subjects has similar content \cite{chaitanya2020contrastive}.
Each volume is grouped into $S$ partitions, and one image sampled from partition $s$ of volume $i$ is denoted as $x_s^i$.
\textit{Local positives} are features of images from the same partition in different volumes.
Given an image $x_s^i$, its feature $q_s^i$ and corresponding positives $P(q_s^i)=\{{k_s^i}^{+}, {k_s^j}^{+}\}$ are formed as follows.
Two transformations (e.g. cropping) are applied to $x_s^i$ to get $\tilde{x}_s^i$ and $\hat{x}_s^i$, which are then fed into 
the main encoder and momentum encoder to generate two representation vectors $q_s^i$ and ${k_s^i}^{+}$, respectively.
Then another image $x_s^j$ is sampled from partition $s$ of volume $j$, and its features $q_s^j$ and ${k_s^j}^{+}$ are generated accordingly.
In this way, the local positives for both $q_s^i$ and $q_s^j$ are formed as $P(q_s^i)=P(q_s^j)=\{{k_s^i}^{+}, {k_s^j}^{+}\}$.

\textbf{Loss function for local positives.}
By using the sampled memory bank $Q^{\prime}$ consisting of both \textit{local} negatives and \textit{remote} negatives,
one local feature $q$ is compared with its local positives $P(q)$ and each negative in $Q^{\prime}$. 
The contrastive loss is defined as:
\begin{equation}\label{equ:loss_memorybank}
\mathcal{L}_{local}=\ell_{q,P(q),Q^{\prime}} = - \frac{1}{|P(q)|} \sum_{k^{+} \in P(q)}  \log \frac{\exp (q \cdot k^{+} / \tau)}{\exp (q \cdot k^{+} / \tau) + \sum_{n\in Q^{\prime}} \exp (q \cdot n / \tau)}.
\end{equation}
where $\tau$ is the temperature and the operator $\cdot$ is the dot product between two vectors.
By minimizing the loss, the distance between $q$ and each local positive is minimized, and the distance between $q$ and each negative in $Q^{\prime}$ is maximized.

\subsection{Global Structural Matching}
\textbf{Remote positives.} We use the remote positives from the shared features to further improve the learned representations.
On each client, we align the features of one image to the features of images in the same partition from other clients. 
In this way, the features of images in the same partition across clients will be aligned in the feature space and more unified representations can be learned among clients.
To achieve this, for one local feature $q$, in addition to its local positives $P(q)$, we define remote positives $\Lambda(q)$ as features in the sampled memory bank $Q^{\prime}$ which are in the same partition as $q$.
\begin{equation}\label{equ:}
    \Lambda(q) = \{ p \ |\ p \in Q^{\prime}, partition(p)=partition(q) \}.
\end{equation}
$partition(\cdot)$ is the partition number of one feature
and $Q^{\prime}$ is defined in Eq.(\ref{equ:contrastive_negatives}).

\textbf{Final loss function.}
By replacing local positives $P(q)$ in Eq.(\ref{equ:loss_memorybank}) with remote positives $\Lambda(q)$  as $\mathcal{L}_{remote}$,
the final loss function for one feature $q$ is defined as:
\begin{equation}\label{equ:}
    \mathcal{L}_{q} = \mathcal{L}_{remote} + \mathcal{L}_{local} = \ell_{q,\Lambda(q),Q^{\prime}} + \ell_{q,P(q),Q^{\prime}}.
\end{equation}
With $\mathcal{L}_{q}$, the loss for one batch of images is defined as $\mathcal{L}_{B}=\frac{1}{|B|}\sum_{q\in B} \mathcal{L}_{q}$, where $B$ is the set of features generated by the encoder from the batch of images.

\section{Experiments}

\textbf{Dataset and preprocessing.}
We evaluate the proposed approaches on the ACDC MICCAI 2017 challenge dataset \cite{bernard2018deep}, which has 100 patients with 3D cardiac MRI images.
Each patient has about 15 volumes covering a full cardiac cycle, and only volumes for the end-diastolic and end-systolic phases are annotated by experts for three structures, including left ventricle, myocardium, and right ventricle. Details of preprocessing can be found in the supplementary material.

\noindent
\textbf{Federated and training setting.}
Following \cite{zhao2018federated}, we use 10 clients. 
We randomly split 100 patients in ACDC dataset into 10 partitions, each with 10 patients.
Then each client is assigned one partition with 10 patients.
We use the proposed FCL approaches to pre-train the U-Net encoder on the assigned dataset partition on each client without labels.
Then the pre-trained encoder (i.e. the final global encoder after pre-training) is used as the initialization for fine-tuning the U-Net segmentation model by using a small number of labeled samples.
The U-Net model follows the standard 2D U-Net architecture \cite{ronneberger2015u} with the initial number of channels set to 48.
We evaluate with two settings for fine-tuning: \textit{local fine-tuning} and \textit{federated fine-tuning}.
In local fine-tuning, each client fine-tunes the model on its local annotated data.
In federated fine-tuning, all clients collaboratively fine-tune the model by supervised FL with a small number of annotations.
Training details can be found in the supplementary material.

\noindent
\textbf{Evaluation.}
During fine-tuning, we use 5-fold cross validation to evaluate the segmentation performance. In each fold, 10 patients on one client are split into a training set of 8 patients and a validation set of 2 patients. 
For each fold, we fine-tune with annotations from $N\in \{1,2,4,8\}$ patients in the training set, and validate on the validation set of the same fold on all clients (i.e. 20 patients). 
Dice similarity coefficient (DSC) is used as the metric for evaluation.

\noindent
\textbf{Baselines.}
We compare the proposed approaches with multiple baselines.
\textit{Random init} fine-tunes the model from random initialization.
\textit{Local CL} performs contrastive learning on each client by the SOTA approach \cite{chaitanya2020contrastive} with unlabeled data for pre-training the encoder  before fine-tuning.
\textit{Rotation} \cite{gidaris2018unsupervised} is a self-supervised pre-training approach by predicting the image rotations.
\textit{SimCLR} \cite{chen2020simple} and \textit{SwAV} \cite{caron2020unsupervised} are the SOTA contrastive learning approaches for pre-training.
We combine these three self-supervised approaches with \textit{FedAvg} \cite{mcmahan2017communication} as their federated variants \textit{FedRotation}, \textit{FedSimCLR}, and \textit{FedSwAV} for pre-training the encoder.
\textit{FedCA} is the SOTA federated unsupervised learning approach for pre-training \cite{zhang2020federated}.

\subsection{Results of local fine-tuning}

\begin{table}[!htb]
	\centering
	\caption{Comparison of the proposed approaches and baselines on \textbf{local fine-tuning} with limited annotations on the ACDC dataset. $N$ is the number of annotated patients for fine-tuning on each client. 
	The average dice score and standard deviation across 10 clients are reported, in which on each client the dice score is averaged on 5-fold cross validation. 
	The proposed approaches substantially outperform all the baselines with different numbers of annotations. 
	}
	\label{tab:exp_local_finetune}
	\resizebox{\columnwidth}{!}{
    \begin{tabular}{lcccc}
    \toprule
    Methods     & $N$=1    & $N$=2    & $N$=4    & $N$=8    \\ \midrule
    Random init & 0.280 $\pm$ 0.037 & 0.414 $\pm$ 0.070 & 0.618 $\pm$ 0.026 & 0.766 $\pm$ 0.027 \\
    Local CL \cite{chaitanya2020contrastive}   & 0.320 $\pm$ 0.106 & 0.456 $\pm$ 0.095 & 0.637 $\pm$ 0.043 & 0.770 $\pm$ 0.029 \\
    FedRotation \cite{gidaris2018unsupervised} & \underline{0.357} $\pm$ 0.058 & \underline{0.508} $\pm$ 0.054 & \underline{0.660} $\pm$ 0.021 & \underline{0.783} $\pm$ 0.029\\
    FedSimCLR \cite{chen2020simple}  & 0.288 $\pm$ 0.049 & 0.435 $\pm$ 0.046 & 0.619 $\pm$ 0.032 & 0.765 $\pm$ 0.033 \\
    FedSwAV \cite{caron2020unsupervised}    & 0.323 $\pm$ 0.066 & 0.480 $\pm$ 0.067 & 0.659 $\pm$ 0.019 & 0.782 $\pm$ 0.030 \\
    FedCA \cite{zhang2020federated}      & 0.280 $\pm$ 0.047 & 0.417 $\pm$ 0.042 & 0.610 $\pm$ 0.030 & 0.766 $\pm$ 0.029 \\
    Proposed    & \textbf{0.506} $\pm$ 0.056 & \textbf{0.631} $\pm$ 0.051 & \textbf{0.745} $\pm$ 0.017 & \textbf{0.824} $\pm$ 0.025 \\ \bottomrule
    \end{tabular}
    }
\end{table}

We evaluate the performance of the proposed approaches by fine-tuning locally on each client with limited annotations.
As shown in Table \ref{tab:exp_local_finetune}, the proposed approaches substantially outperform the baselines.
First, with 1, 2, 4, or 8 annotated patients, the proposed approaches 
outperform the best-performing baseline by 0.149, 0.123, 0.085, and 0.041 dice score, respectively.
Second, the proposed approaches significantly improve the annotation efficiency.
For example, with 1 annotated patient, the proposed approaches achieve a similar dice score to the best-performing baseline with 2 annotations (0.506 vs. 0.508), which improves labeling-efficiency by 2$\times$.

\subsection{Results of federated fine-tuning}

\begin{table}[!htb]
	\centering
	\caption{Comparison of the proposed approaches and baselines on \textbf{federated fine-tuning} with limited annotations on the ACDC dataset. $N$ is the number of annotated patients for fine-tuning on each client. 
	The proposed approaches significantly outperform all the baselines with different numbers of annotations.
}
	\label{tab:exp_federated_finetune}
	\resizebox{\columnwidth}{!}{
    \begin{tabular}{lcccc}
    \toprule
    Methods     & $N$=1    & $N$=2    & $N$=4    & $N$=8    \\ \midrule
    Random init & 0.445 $\pm$ 0.012 & 0.572 $\pm$ 0.061 & 0.764 $\pm$ 0.017 & 0.834 $\pm$ 0.011 \\
    Local CL \cite{chaitanya2020contrastive}   & 0.473 $\pm$ 0.013 & \underline{0.717} $\pm$ 0.024 & 0.784 $\pm$ 0.015 & 0.847 $\pm$ 0.009 \\
    FedRotation \cite{gidaris2018unsupervised} & \underline{0.516} $\pm$ 0.015 & 0.627 $\pm$ 0.074 & \underline{0.821} $\pm$ 0.015 & \underline{0.867} $\pm$ 0.010 \\
    FedSimCLR \cite{chen2020simple}  & 0.395 $\pm$ 0.023 & 0.576 $\pm$ 0.046 & 0.788 $\pm$ 0.014 & 0.859 $\pm$ 0.011 \\
    FedSwAV \cite{caron2020unsupervised}    & 0.500 $\pm$ 0.015 & 0.594 $\pm$ 0.058 & 0.815 $\pm$ 0.015 & 0.862 $\pm$ 0.010 \\
    FedCA \cite{zhang2020federated}      & 0.397 $\pm$ 0.020 & 0.561 $\pm$ 0.047 & 0.784 $\pm$ 0.015 & 0.858 $\pm$ 0.011 \\
    Proposed    & \textbf{0.646} $\pm$ 0.052 & \textbf{0.824} $\pm$ 0.004 & \textbf{0.871} $\pm$ 0.007 & \textbf{0.894} $\pm$ 0.006 \\ \bottomrule
    \end{tabular}
    }
\end{table}

We evaluate the performance of the proposed approaches by collaborative federated fine-tuning with limited annotations.
Similar to local fine-tuning, the proposed approaches significantly outperform the SOTA techniques as shown in Table \ref{tab:exp_federated_finetune}. 
First, with 1, 2, 4, or 8 annotated patients per client, 
the proposed approaches outperform the best-performing baselines by 0.130, 0.107, 0.050, and 0.027 dice score, respectively.
Second, the proposed approaches effectively reduce the annotations needed for fine-tuning. For example, with 2 or 4 annotated patients per client, the proposed approaches achieve better performance than the best-performing baseline with 2$\times$ annotated patients per client,
which achieve more than 2$\times$ labeling-efficiency.
Third, compared with local fine-tuning in Table \ref{tab:exp_local_finetune}, all the approaches achieve a higher dice score. 

\section{Conclusion and Future Work}
This work aims to enable federated contrastive learning for volumetric medical image segmentation with limited annotations.
Clients first learn a shared encoder on distributed unlabeled data and then a model is fine-tuned on annotated data.
Feature exchange is proposed to improve data diversity for contrastive learning while avoiding sharing raw data.
Global structural matching is developed to learn an encoder with unified representations among clients.
The experimental results show significantly improved segmentation performance and labeling-efficiency compared with state-of-the-art techniques.

\noindent
\textbf{Discussion.}
Sharing features needs additional communication, and we will explore techniques to reduce the communication cost.
Besides, we will explore defenses such as \cite{li2019deepobfuscator} against inversion attacks \cite{lyu2020threats} for improved security.

\bibliographystyle{splncs04}
\bibliography{ref}

\end{document}